\documentclass[aps,showpacs]{revtex4}
\usepackage{fullpage,amsfonts,amssymb,pst-node,epsfig}

\begin{document}
\title{Fano theory for hadronic resonances: \\
the rho meson and the pionic continuum}
\author{N.E. Ligterink\thanks{E-mail: nol1@pitt.edu}}
\affiliation{Dept. of Physics and Astronomy, University of
Pittsburgh, 3941 O'Hara Street, Pittsburgh, PA 15260, U.S.A.}
\date{\today}
\begin{abstract}
We develop a model-independent analysis of hadronic scattering data
in the resonance region, where the
resonance shape follows from the matrix elements of a Hamiltonian.
We investigate the rho meson in the tau decay. We demonstrate that
the rho meson resonance in the two-pion decay of the tau lepton
is described well through the coupling of a bare rho meson
to the two-pion and the four-pion continuum.
Furthermore, this four-pion
continuum corresponds with the data of the four-pion decay channel
of the $\tau^-$ at energies up to $1.1$ GeV.
\end{abstract}

\maketitle

\section{Introduction}

The most obvious unstable hadronic states, or resonances, are single
absorption peaks in the $\pi N$ scattering data in a particular channel, with a
Breit-Wigner, or Lorentz, shape, where the width is related to the 
coupling to mainly pionic decay channels. However, 
resonances generally have distorted shapes due to the 
presence of other resonances or thresholds of competing channels. 
Their shape might look nothing like the Breit-Wigner shape that
is often used to fit the data, and the position of the peak might
shift through inter-resonance and resonance-continuum interactions.
The extraction of single resonances
from complex scattering data is still strongly model dependent~\cite{VDL}.

Today, baryons are no longer tied to $\pi N$ scattering,
but are also observed in other experiments, such as $\gamma N$
scattering. They are particles in their own right, which can be
probed with different experiments. However, hadrons are not elementary
particles.
QCD was discovered as the underlying microscopic field theory
of composite hadrons. In principle hadrons are bound states of
quarks, but deriving their properties directly from QCD has 
proven difficult. 
Therefore, the interest in hadronic resonances
has been extended from basic resonance parameters, such as 
the mass and the
width, to microscopic properties that can test the quality of 
a wide range of microscopic
models of hadrons. Although most models  reproduce the mass spectrum 
with a reasonable accuracy, microscopic properties, such as charge radii and
magnetic moments, and branching ratios of decays are currently under
investigation.

Furthermore, quark models allow for states, hybrids or exotics,
that do not consist of three quarks, or simple quark-anti-quark pairs.
However, to extract a clear signal for these states
from experiment is a formidable task. Therefore, it is important to
understand the conventional hadronic states and their coupling to
possible exotic states better.

Since the first analysis of hadronic resonances there have been many
theoretical developments. Quantum field theory has replaced quantum
mechanics as the fundamental theory for small-scale physics, and renormalization
is an unfortunate, but integral part of a modern approach based on local
interactions. In hadron physics the issue of handling divergences and
renormalization is often by-passed by introducing form factors or
vertex functions.
Vertex form factors which smear the interactions and make
further regularization superfluous fail to have universal applicability;
each experiment within a particular energy range
is explained with a model which has its own form factors.
Global models that fit the world data do a good job in the energy
range of the world data, but these models would not necessarily
lead to good predictions for new results at higher energies.

The local interactions in field theory seem essential to maintain 
relativistic covariance, and, although models with vertex form factors
can reproduce the data well with a limited number of parameters, they 
are tied to a particular experiment, a
particular energy range, and a particular set of channels, and they say
very little general about a particular hadronic state. Moreover, a different
set of vertex form factors might do just as well.

Furthermore, there are bounds and constraints, such as
unitarity and analyticity, on the models that describe 
scattering. They were put on a firm footing by study of the analytical 
properties of the S-matrix, and are incorporated in quantum field
theory. However, since hadrons are not elementary particles, there
is no fundamental field theory, with a small number of parameters,
that describes all of hadron dynamics. Model field theories of hadron physics 
serve to bring order in the complexity, to restrict the model space,
and to implement constraints from symmetries. 

Separate from the choice of the model hadronic field theory, 
the theory of resonances has unfortunately not kept pace with the 
developments in field theory in general. Consistency requires that 
the imaginary  part of the mass 
is of an unstable particle follows from a microscopic model that links
the particle with its decay channels. Some formal developments have 
been made~\cite{MS59,Zwa63}; however, in practice the models underlying 
the resonance
shapes that describe actual data, which can be very complicated, are
in essence the same as those of thirty years ago~\cite{BW52,tola}. 
These models aim to locate the
resonance poles, but gives very little other information that can
constrain models of hadron interactions. Microscopic properties
of hadrons remain hidden in the data.

Some theoretical progress has been made.
For example, resonances have been studied and renormalization is
handled in chiral perturbation theory~\cite{KKW}.
However, the typical energy of hadronic
resonances is too high and lies beyond the region of applicability of 
chiral perturbation theory. At these energies
the chiral symmetry is not important, and other features, such as
form factors or a large number of low-energy parameters dominate 
the fit to the data. The hadronic Lagrangian is not
restricted by chiral symmetry at these energies, and many constants 
need to be fitted to the data.

In most cases, attempts to describe medium-energy hadronic physics
from an underlying field theory start with a chiral Lagrangian
which is augmented with additional constraints. For example,
one important, but approximate conservation law in medium-energy 
hadronic physics, which has been utilized, follows from vector 
meson dominance.  This
suggests there is a universal hadronic current by which the
photon and the rho meson couple to all hadrons~\cite{HLS}.
Furthermore, the large $N_c$ limit has some implications for
amplitudes, which can restrict the $SU(3)_f$ chiral Lagrangian
so that it has predictive power~\cite{Lut00}.
In more traditional chiral Lagrangian approaches unitarity has
been used to improve the predictions beyond the original
range of applicability~\cite{oller}.
Finally, from a more traditional hadronic few-body dynamics point 
of view, a consistent implementation of covariance can extend
the range of validity and restrict the model dependence~\cite{Pic01}.
For most of these approaches, scales and form factors still need to be
introduced. Furthermore, the models are more designed to test the assumptions
made, than to extract universal quantities from a wide range of
scattering data. In all these approaches hadrons enter as the
elementary fields. Independently, within
the Dyson-Schwinger framework, there are investigations on the way to
use their rho meson Bethe-Salpeter results in baryon and meson 
form factors~\cite{tandy}, where the composite nature of hadrons in
terms of quarks is resolved.

Our aims are much simpler; we would like to determine universal quantities,
like the matrix elements of a Hamiltonian, directly and consistently
from the data.
Consistent tools for the analysis of resonances that satisfy
unitarity, analyticity are developed. They meet a microscopic model for
hadrons halfway in terms of matrix elements and bare masses.
The divergences that arise from the local
interactions are renormalized, and the only parameters present are 
a few coupling constants and bare masses. 
Simple field-theoretical results, such as the bubble summation for a 
single channel are reproduced. The method is
much more versatile, and this versatility is paramount. 
The coupled channel data should yield an Hamiltonian with the same
asymptotic states first.  If there is an important inelastic
cross section, the only models that can be trusted are those which
explicitly take into account these channels. Possible candidates 
for resonances can be added and their effects on the data studied. It
is not a matter of explaining a particular set of data, but
it is important to have independent verifications of resonances
in independent experiments. 

Any other input is considered part of a microscopic model, and should be 
kept in its simplest form. The approach does not hinge
on a particular approximation, but solely on one condition. 
Any approximation should be made at the level of the Hamiltonian
and it should maintain Hermiticity.
The Hamiltonian is a good starting point and end point of such tools:
formulate a model in terms of a Hamiltonian, and fix the matrix elements
with the data. Unitarity, analyticity, and other constraints on the data
are automatically fulfilled within this approach. 

This method for analysis of hadronic
resonances is designed for the case of a strong coupling between continua and
multiple resonance states. This paper is the first step in developing a set of
tools for a fully unitary approach to coupled-channel scattering
problems, which is under investigation at the moment. Furthermore, 
it establishes contact with the local field theories that should describe these
systems, and renormalization is handled.

In most studies resonances are fitted with Breit-Wigner resonance shapes,
however, resonances have often more complicated shapes. 
In atomic theory there is
a wide range of methods developed that describe resonances or
auto-ionization profiles. In hadronic physics the interaction is
stronger. The resonances have generally a larger width, and 
therefore more complicated shapes as they interact with other
resonances and the asymtotic states. 
The method used here is an adaption of a method developed by
Fano in the 1960's.

Fano developed~\cite{Fan61,Cow82,BR97} an analysis 
for atomic resonances by writing
down a general Hamiltonian for such systems. This Hamiltonian
can be diagonalised exactly, thereby yielding direct relations
between scattering data and the parameters of the Hamiltonian.
It is based on the separation of the problem into two parts:
the continuum, characterized by the energy of the asymptotic states
in a particular channel, and resonances, which in the absence of
a continuum are discrete eigenstates, such as, the states
that would follow from a constituent quark model, large $N_c$
approximation, or a quenched lattice calculation.

The continuum should be considered an
eigenstate  of the decoupled system as well. The wave function could 
be deformed at the
origin with respect to the non-interacting wave function due to
direct continuum-continuum interaction as would occur in field theory. 
In this case the continuum can be composite; with asymptotic states 
without fixed particle numbers, as we will see in section~\ref{rho}
where we discuss the rho meson as an example.
By splitting the problem in two and expressing it in the eigenstates of the 
sub-problems many matrix elements are zero due to orthogonality of the
eigenstates of the sub-problems. 

As a first example we analyze the rho meson resonance in the tau
decay. The resonance
shape dominates the hadronic decay of the tau lepton into two
pions. We choose to describe this problem with a bare rho  which
couples to a two-pion state.
The energy dependence of
the coupling can be modeled by a chiral Lagrangian.
To extend this simple model, there are two main avenues: introduce
a more complicated energy dependence, motivated by some
microscopic model, or introduce the next state that is
important, which is the four-pion state (with a threshold at 
$558$ MeV). For the data it is clear that the four-pion decay 
of the tau lepton dominates the two-pion decay above 1 GeV.
The effects of four-pion states are
hard to handle in chiral perturbation theory, since this
involves at least a three-loop diagram and many intermediate states,
such as $\omega\pi$, $\rho\rho$, $\pi\pi\rho$, and $a_0\pi$
mesons~\cite{BR69}.
However, we will see that a simple four-pion state does precisely
what is needed: it corrects the high-energy part of the two-pion
decay, and gives a quantitative prediction of the four-pion 
decay. In the Hamiltonian approach the details of  
of the rho four-pion coupling cannot be resolved with
present-day data. The two effective coupling constants;
the two-pion and the
four-pion decay of the rho meson are sufficient to describe the
hadronic tau lepton
below $1.1$ GeV. Above this
energy there are many channels which can only be distinguished by
looking at invariant masses of subsystems in multi-pion states to which
most of these hadronic states decay. At the
moment only the three-pion decay of the intermediate omega meson has a
clear signal~\cite{Eck}, mainly because of the narrow width of the omega.

Many theoretical studies of the rho meson focus on combining vector
meson dominance (VMD)~\cite{Sak60} with chiral symmetry.  In these models
the photon
couples to most hadrons through a universal conserved vector current 
(CVC). 
The microscopic foundation of this current, and
its underlying symmetry, is poorly
understood, and a hidden local symmetry (HLS) is
postulated~\cite{HLS}. This symmetry mixes the photon with the rho
meson. The additional knowledge resulting from this assumption has led 
to a reasonable understanding of the shape of the rho meson due to 
coupling with the two-pion continuum.

However, since chiral perturbation theory focusses on low-energy behavior,
the predicted higher energy behavior does not agree with the data,
because the higher order derivative 
couplings violate unitarity at intermediate energies.
Therefore, the high-energy part of the rho resonance fails the
data completely. Improved HLS models
unitarize previous results, and somewhat improve agreement. We, on the
other hand,
find the correct high-energy behavior by simply including the four-pion
continuum.  

In the next section, we will develop the general theory of resonances
given a Hamiltonian with a number of discrete states and a continuum.
In Section \ref{cf} we show that the matrix elements that appear in the
Hamiltonian can be related to Feynman diagrams. For a first application
to hadronic resonances we apply the method to the rho meson in Section
\ref{rho}. In Section \ref{results} we present the results, and the
final section we discuss future applications and improvements of this
method.

\section{Theory}

Let us first consider a general Hamiltonian which couples a continuum to a
set of discrete states that are orthogonal with respect to each other.
Many systems can be cast in this form (it does not allow for
continuum-continuum interactions, which is a topic of current
investigations). The Hamiltonian has the general form:
\begin{eqnarray}
H & = &  \sum_{i=1}^k |i\rangle \epsilon_i \langle i | + \int
 d \Delta \  |\Delta \rangle \Delta \langle \Delta | \nonumber \\
 & & + \sum_{i=1}^k \int W_i(\Delta) d \Delta \
\left( |\Delta \rangle e^{-i\phi_i(\Delta)} \langle i |
\  + \ |i \rangle e^{i\phi_i(\Delta)} \langle \Delta | \right) \ \ ,
\label{ham}
\end{eqnarray}
where $\epsilon_i = \sqrt{{\bf p}^2+M_i^2}$ with ${\bf p}$ the momentum of the system,
and $M_i$ is the mass of the discrete, bare state $i$. The continuum state
$|\Delta \rangle$ is labeled with its energy $\Delta$.  The coupling is
parametrized with the unknown coupling functions $W_i$
and phases
$\phi_i$. In practice, such Hamiltonian are restricted to a particular
channel, with a given partial wave, charge, and other conserved quantum
numbers, which we therefore omit from the start.

All the integrals are principal value integrals, additional
terms that arise from the singularities will be taken in account
explicitly, rather than implicitly in a particular pole or
$i \epsilon$ prescription, which places the pole at a particular
side of the contour integration. The latter is used in scattering theory,
where the pole part is associated with open channels, or asymptotic,
states. Within an eigenfunction interpretation, we do not necessarily
interpret singular parts of the eigenstate with out-states, they
could be in-states as well, or, if the interaction is strong, mix with
the resonance state. This information is carried in the occupation
numbers $\alpha_i$ and $\beta$, which are normalized to unity, even for
a singular functional dependence of the spectroscopic strength
$\beta$ on the energy.

The eigenstates $|\omega\rangle$ with the continuous energy $\omega$ of the
Hamiltonian can be expressed as:
\begin{equation}
|\omega\rangle = \int d \Delta \beta(\omega,\Delta) |\Delta \rangle +
\sum_{i=1}^k \alpha_i(\omega) |i\rangle \ \ .
\end{equation}
where $\alpha_i$'s are the occupation numbers of the discrete state,
and $\beta$ is the spectroscopic amplitude, or occupation density, of
the continuum states.
Substituting this into the Hamiltonian, Eq.~((\ref{ham})),  we can express
$\beta(\omega,\Delta)$ in terms of the $\alpha$'s:
\begin{equation}
\beta(\omega,\Delta) = \left( \frac{1}{\omega-\Delta} + z(\omega) \delta(\omega-\Delta)
\right) \sum_{i=1}^k \alpha_i(\omega) W_i(\Delta) e^{-i\phi_i(\Delta)} \ \ .
\label{beta}
\end{equation}
where $z(\omega)$ can be any regular function to be determined later.
$z(\omega)(\omega-\Delta)\delta(\omega-\Delta)$ vanishes.
Substituting $\beta$ back into the second equation of the Hamiltonian, 
Eq.~(\ref{ham}), yields:
\begin{eqnarray}
(\omega -\epsilon){\bf \alpha}(\omega) & = &   \pi {\cal F}(\omega) \cdot
 {\bf \alpha}(\omega)+
z(\omega)  {\bf F}(\omega) \cdot  {\bf \alpha}(\omega)  \ \ ,
\label{mat}
\end{eqnarray}
where $(\omega -\epsilon)$ is a diagonal matrix with entries $\omega - \epsilon_i$.
The hermitian matrix ${\bf F}$, and its Hilbert transform ${\cal F}$
necessary for the consistency condition on $z(\omega)$ are defined:
\begin{eqnarray}
F_{ji}(\xi) & = &  W_i(\xi) W_j(\xi) e^{i(\phi_j(\xi) -\phi_i(\xi))}\ \ , \\
{\cal F}_{ji}(\eta) & = & \frac{1}{\pi} \int d \xi \frac{F_{ij}(\xi)}{\eta-\xi} \ \ .
\label{real_F}
\end{eqnarray}
Eq.~(\ref{mat}) is multiplied from the left by:
\begin{equation}
 {\bf W}^\dagger(\omega) \cdot\left((\omega -\epsilon)- \pi {\cal  F}(\omega) \right)^{-1}  \ \ ,
\end{equation}
where the vector
${\bf W}^\dagger(\omega) = (W_1(\omega) e^{-i\phi_1(\omega)}, W_2(\omega) e^{-i\phi_2(\omega)},
\cdots, W_k(\omega) e^{-i\phi_k(\omega)} )$.
The whole expression contains a common factor 
${\bf W}^\dagger(\omega) \cdot \alpha(\omega)$, which can be
divided out, which yields a closed form expression for $z(\omega)$:
\begin{equation}
z(\omega) = \left(  {\bf W}^\dagger(\omega) \cdot\left((\omega -\epsilon)-
\pi {\cal  F}(\omega) \right)^{-1} \cdot {\bf W}(\omega) \right) ^{-1} \ \ .
\label{z}
\end{equation}
At this point it is important to realize the significance of $z(\omega)$
above:
it is the projected resolvent $((\omega -\epsilon)-\pi {\cal
F}(\omega))^{-1}$ of the discrete spectrum, perturbed through the coupling with the
continuum.  Given that ${\cal F}$ is hermitian, the matrix $(\epsilon - \pi {\cal F})$ will
have a discrete spectrum $\omega_1,\omega_2,\cdots, \omega_k$,
\begin{equation}
\Omega =
\left(\begin{array}{cccc} \omega_1 & & & \cr
                     & \omega_2 & & \cr
                     & & \ddots & \cr
                     & & & \omega_k \end{array} \right) =
U^\dagger(\omega) \cdot \left( \epsilon - \pi {\cal F}(\omega) \right) \cdot U(\omega) \ \ .
\end{equation}
In order to simplify the results, it is convenient to work in the basis of the true spectrum
of the discrete system. In this case the inverse, that appears in Eq.~(\ref{z}), is trivial. We
perform the substitutions:
\begin{eqnarray}
\alpha(\omega) & = & U \cdot {\bf a}(\omega) \ \ , \\
{\bf W}(\omega) & = &  U \cdot {\bf V}(\omega) \ \  ,
\end{eqnarray}
and likewise for the hermitian conjugates.

The normalization of the state $|\omega \rangle $ yields:
\begin{eqnarray}
\langle \omega' | \omega \rangle  & = &
 \alpha^\dagger(\omega') \cdot \alpha(\omega) \\
& &  +  \alpha^\dagger(\omega')\cdot \frac{z(\omega') {\bf W}(\omega') {\bf W}^\dagger(\omega')  -
\pi {\cal F}(\omega')}{\omega'-\omega} \cdot \alpha(\omega)  \nonumber \\
& & - \alpha^\dagger(\omega')\cdot \frac{z(\omega) {\bf W}(\omega) {\bf W}^\dagger(\omega)  -
 \pi {\cal F}(\omega)}{\omega'-\omega} \cdot \alpha(\omega)\nonumber  \\
& &  + (\pi^2 + z^2(\omega))\delta(\omega-\omega')
\alpha^\dagger(\omega')\cdot {\bf W}(\omega') {\bf W}^\dagger(\omega)
\cdot \alpha(\omega)  \ \ , \nonumber
\end{eqnarray}
where the $\pi^2$ contribution results from the pole term of the product of the two principal
value singularities in the spectroscopic densities $\beta$.
If we substitute the new variables, {\bf a} and {\bf V},
it is easy to see that the first three terms cancel each other.
The state $|\omega \rangle $ is an eigenstate by construction and: 
\begin{equation}
\langle \omega' | \omega \rangle  =  (\pi^2 + z^2(\omega))\delta(\omega-\omega')\alpha^\dagger(\omega')\cdot {\bf W}(\omega') {\bf W}^\dagger(\omega)
\cdot \alpha(\omega) \ \ .
\end{equation}
The probability of finding the system in the discrete state $|i\rangle$, given it was in
the discrete state $|j\rangle = |\psi(t=0)\rangle$ is:
\begin{equation}
\langle i |\psi(t)\rangle = \int d \omega \ \alpha_i (\omega) \alpha_j^\ast(\omega) e^{-i \omega t}   \ \ ,
\label{decay}
\end{equation}
where the occupation numbers are given by:
\begin{equation}
\alpha_i(\omega) =
\frac{W_i(\omega) e^{- i \phi_i(\omega)}}
{{\bf W}^\dagger {\bf W}\left[z(\omega) + i \pi\right]} \ \ ,
\end{equation}
and we have chosen a convenient phase. At the position of a
resonance $\omega = \omega_j$, $z(\omega_j)$ vanishes, however, this
is not necessarily the resonance maximum, as the
coupling function depends on the energy. If the coupling
function is a constant, a single resonance shape reduces to a Breit-Wigner
form:
\begin{equation}
\alpha_j(\omega) = \frac{W_j}{ (\omega - \omega_j) + i \pi W_j^2 } \ \ .
\end{equation}

Likewise, the phase shift follows from the inspection of Eq.~(\ref{beta})
for the spectral density of the continuum state. The principal value
singularity yields the scattering wave function proportional to 
$-\pi \sin (k(\omega) + \delta_l)$, while the delta function yields
the scattering wave function proportional to $\sin (k(\omega) + \delta_l
+\frac{\pi}{2})$.
Therefore, their relative strength $\pi/z(\omega)$  determines the
effect of the coupling to the discrete states to the phase shift
$\delta_r$, which yields
\begin{equation}
\delta_r = - \arctan \frac{\pi}{z(\omega)} \ \ .
\label{phase}
\end{equation}

Clearly, once the eigenstate is known, in terms of the occupation numbers $\alpha_i$ of the
discrete states and the occupation number $\beta$ of the continuum, at every energy we can 
determine any quantity we like. The decay of a state is given by the probability of
finding the system in that state, Eq.~(\ref{decay}). In a scattering process, 
the incoming state
must be projected on the continuum part of the eigenstate, and the scattering continuum
state can be reached subsequently via one of the discrete states.
Transition amplitudes are therefore proportional to $W_j \alpha_j$. For
a single channel the $T$-matrix is:
\begin{equation}
T = \frac{1}{z(\omega) + i \pi} \ \ ,
\end{equation} 
which preserves unitarity.

\section{Coupling functions}
\label{cf}

\begin{figure}
\centerline{\includegraphics[width=4cm]{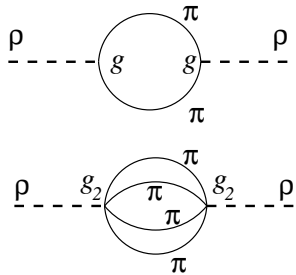}}
\caption{The two diagrams from which the coupling functions $W$ and $W_2$, between the
discrete state and the continuum, is derived.}
\label{fig2}
\end{figure}

The masses $M_i$ and the coupling functions $W_i$ are to be determined by other means.
We assume that the masses follow from some QCD or constituent-quark
based model. Although, the coupling functions; the pion-hadron interaction, could
in principle be determined by the same means, in practice that
turned out to be a formidable task.  Therefore, we turn to Lagrangians with 
explicit pions and chiral symmetry for
the coupling functions~\cite{BR69}. The easiest connection is made
midway. Although the meaning of the Green's function, or resolvent, is
different in the study of eigenstates, which is stationary, and the
study of scattering, which is dynamical, the actual mathematical form
is the same. At many levels one can find relations between expressions
in scattering theory and stationary eigenvalue problems. We choose the
equivalence that is most convenient to us, which is the equivalence
of the one-particle self-energy. With 
chiral Lagrangians
we can determine the one-particle irreducible
self-energy correction $\Sigma$ of hadrons 
due to coupling to the pion field, 
which is given in terms of Feynman diagrams. We can equate this with the 
real part of energy shift due to the coupling of the $i$-th discrete
state with the continuum $\pi {\cal F}_{ii}$
in the Fano theory, which also yields the real part of the self-energy
due to the coupling with the continuum:
\begin{equation}
\int_{PV} d \Delta \frac{W_i^2(\Delta)}{\omega - \Delta} \equiv
\Sigma_i(\omega) =  \frac{1}{2 \omega}  {\rm Re} \int d^{4n} k\  I(k) \ \ ,
\label{eqW}
\end{equation}
where $I$ is the integrand of the truncated Feynman self-energy diagram $\Gamma$ 
for the bare hadron $i$. 
Hence $W_i^2$ is the phase-space of the continuum times the appropriate factors:
\begin{equation}
W_i^2(\omega) =  \frac{\pi}{2 \omega} {\rm Im} \int d^{4n} k\  I(k) \ \ ,
\end{equation}
where $\omega$ is the sum of the free energies of the particles in the 
continuum state, which, in a Feynman diagram with energy conservation at
the vertices, equals the total incoming energy.
Similarly, in the right-hand side of Eq.~(\ref{eqW}) we can
recognize the corresponding Goldstone diagram~\cite{BR86}.

This connection between the coupling function $W$ and the imaginary part 
of the Feynman diagram
guarantees that the tree-level and bubble-sum results in this approach
and with Feynman diagrams are the same. 
However, the Hamiltonian description
does not automatically include the backward, or anti-particle,
contributions, since these are related to separate states. These
states must be added explicitly. The major advantage is that the 
resulting linear eigenvalue equation
in the energy can be solved exactly, even in the case of 
multiple resonances and multiple decay channels. In the language of
Feynman diagrams this statement corresponds to the simultaneous bubble sum of
the different intermediate states.

\begin{figure}
\centerline{\includegraphics[width=6cm]{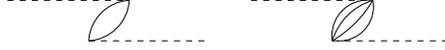}}
\caption{The two backward, or Z diagrams. The inclusion of these two
states in the coupling functions restores covariance, but does not
affect the shape of the resonance, only coupling constants change slightly.}
\label{fig3}
\end{figure}

For the rho meson, analyzed below, we included the anti-particle states,
that is, we combined $\rho\to\pi\pi\to\rho$ with the
$\rho\to\rho\rho\bar\pi\bar\pi\to\rho$ process and  
 $\rho\to\pi\pi\pi\pi\to\rho$ with
$\rho\to\rho\rho\bar\pi\bar\pi\bar\pi\bar\pi\to\rho$. (See Fig.
\ref{fig3}.) Although
this restores covariance it has little effect on the actual resonance
shape, since the threshold of the $\rho\rho\pi\pi$ state
is already above $1800$ MeV. The calculation of the self-energy
simplifies for the rho meson; the functions depend on $s = \omega^2,
\Delta^2$ only, since the forward and the backward contributions combine.
 Schematically:
\begin{equation}
\int d \Delta \frac{f(\Delta^2)}{\omega - \Delta} + \int d \Delta
\frac{f(\Delta^2)}{\omega - (2 \omega + \Delta)} =
\int d \Delta^2 \frac{f(\Delta^2)}{\omega^2 - \Delta^2} \ \ .
\label{ap}
\end{equation}

The self-energy Eq.~(\ref{eqW}) is highly divergent for many pionic problems. 
We will use on-shell
subtraction as our renormalization procedure, which requires implicit 
counterterms such
that the masses $M_i$ and the wave function normalization 
$\langle i |  j \rangle = \delta_{ij}$ 
remain unaltered. In practice the wave-function renormalization means
a unit residue of the resonance pole. Below we discuss the rho meson, 
and show explicitly how these problems are handled.

\section{The rho meson}
\label{rho}

The rho meson plays an important role in photon-hadron interactions at 
intermediate energies.
It seems that the photon couples to hadrons mainly via the isovector-vector 
rho meson, which in its
turn, couples to a universal hadronic current via a coupling constant $g$.
We see the rho meson mainly through its decay to two pions. So in order 
to describe the physical
rho meson, we assume a discrete state $|\rho \rangle$ at rest, which couples 
to the two-pion continuum 
$|\pi\pi(\Delta) \rangle $ with energy $\Delta$.
We have the following matrix elements:
\begin{eqnarray}
\langle \rho |H| \rho \rangle  & = & M  \ \ ,\\
| \langle \rho | H | \pi\pi (\Delta) \rangle |^2  & = &  
\frac{g^2 (\Delta^2 - 4 m_\pi^2)^\frac{3}{2} \theta(\Delta-2 m_\pi)}{
48 \pi \Delta}  = W^2(\Delta) \ \ ,\\
\langle\pi\pi (\Delta') | H | \pi\pi (\Delta) \rangle & = & \Delta \delta(\Delta - \Delta') \ \ ,
\end{eqnarray}
where the coupling function follows from the interaction term 
$ g \vec \rho^\mu \cdot [\vec \pi \times \partial_\mu \vec \pi] $ in the chiral
Lagrangian, and the vectors are in the isospin space.
In practice this
particular coupling is only important for the right threshold behavior,
for large energies the results are dominated by the
phase space.  
For the moment we will assume that there is no self-interaction in
the two-pion continuum, therefore the coupling function $W$ follows the
phase space. The explicit unitarity of the approach and the on-shell
renormalization lead to the correct large energy behavior.

In the future we will investigate the effect of the
continuum-continuum interaction on the results.
We also include the next 
contribution to the physical rho, which, considering
the possible states in this channel and their energies, must be the
four-pion state, with a threshold at $558$ MeV.
From an energy of about $1$ GeV there are many other intermediate 
decay channels
important, e.g.: $\pi^6$, $\omega \pi$, $\rho\pi^2$
and $a_0\pi$ between $900$ MeV and $1.5$ GeV~\cite{BR69}. 
They are to be taken in account if one wants to
analyze rho decay at energies above $1$ GeV. 
For example, the $\pi\omega$ system has a threshold
($921$ MeV) close enough to the $\rho$ mass ($770$ MeV) to yield
contributions that do not simply follow the four-pion phase space.
However, we leave this to future investigations. Although channels open
at a slightly lower energy than $1.1$ GeV, their effect is not strong at threshold
due to the low-energy behavior manifest in the derivative couplings of
the chiral Lagrangian. 

From the chiral Lagrangian there are many intermediate states that lead to the four-pion
state from the discrete rho, however, all these states are highly virtual, 
so we can approximate
the coupling between the discrete rho and the four-pion 
state with a single derivative coupling. (See Fig.~\ref{fig2}.) This
yields a complicated integral, equivalent to a three-loop calculation in
Feynman perturbation theory.
Ignoring these correlations in the four-pion state, the additional matrix elements are
well-approximated by:
\begin{eqnarray}
| \langle \rho | H |\pi\pi \pi\pi (\Delta)\rangle  |^2 & = & 
\frac{g_2^2 (\Delta^2 - 16 m_\pi^2)^\frac{9}{2}\theta(\Delta-4 m_\pi)}{
M^5 \Delta^3}  \equiv W_2^2(\Delta)  \ \ , \\
\langle\pi\pi\pi\pi (\Delta') |H | \pi\pi\pi\pi (\Delta) \rangle & = & \Delta \delta(\Delta - \Delta') \ \ .
\end{eqnarray}
From semi-analytical three-loop calculations~\cite{Lig00} one can see that 
the approximation
for $W_2$ is reasonable, and the eigenstates of the complete system turn
out to be stable under small variations in the coupling function $W_2$.
The continuum is composite.
It contains a two-pion and a four-pion fraction to the ratio $W/W_2$. We did not allow a
direct coupling between them, so they mix with ratios proportional to the coupling 
strengths to the discrete state $|\rho \rangle$.

The real parts, from Eq.~(\ref{real_F}), are divergent integrals, as
expected for local interactions. Therefore we need to regularize
the results and state how we determine the finite parts.
In hadronic physics renormalization has to be pragmatic, since
most theories for hadronic physics are not renormalizable; there
is no fixed set of parameters, masses and coupling constants that
can re-absorb all the divergences by a proper redefinition. Therefore,
at any order or scale a number of observables are needed to fix
the unknown finite renormalizations of the system. In principle a lot
of fine-tuning, to fit theory to experiment, can be done here. However, 
we take the view that this fine-tuning means fitting the short and
medium range physics, which should be done with a proper microscopic
theory of hadrons. The bare rho is the result of miscroscopic
interaction that should incorporate the short range effects of
pions.
Therefore all the finite renormalizations are chosen such that the
contributions vanishes at the rho mass $M$. This is a
choice, but it separates analysis of scattering
data from any model assumptions. If some additional knowledge
about the structure of the rho meson exists, it should end up in
the coupling function. This is important to make the analysis based
on the Hamiltonian scale independent, since scale dependence implies model 
dependence.

For this purpose we use on-shell renormalization; subtracting
the low-order divergent terms in Taylor expansion in 
$(\omega^2-M^2)$ of the self-energy ${\cal F}$. Note that 
since we have included the anti-particle diagrams, Eq.~(\ref{ap}), our
self-energy ${\cal F}$ is a function of $s=\omega^2$ rather than
$\omega$, hence the expansion is in $(\omega^2-M^2)$ rather than
$(\omega - M)$. Odd terms are absent and do not require
renormalization. 

The two-pion coupling function leads to the ordinary mass and
wave-function renormalization $c_0$ and $c_1 (\omega^2-M^2)$.
The four-pion coupling function requires subtractions
$c^{(2)}_0$,  $c^{(2)}_1 (\omega^2-M^2)$, $c^{(2)}_2 (\omega^2-M^2)^2$,
and $c^{(2)}_3 (\omega^2-M^2)^3$, which are set identical to
the same divergent terms in ${\cal F}_{\pi^4}$: 
\begin{equation}
c^{(2)}_n = \frac{1}{n!} \left. \frac{\partial^n {\cal
F}_{\pi^4}(\omega^2)}{ \partial
(\omega^2)^n} \right|_{\omega = M} \ \ .
\end{equation} 
The effects of the self-energy are small over the whole range of the
rho resonance shape, from the two-pion threshold to $1.2$ GeV,
which yields effects of the order of a couple of percent.

Since the finite renormalization  is model independent there are only 
two parameters to fit the data. These are the two coupling constants $g$ and $g_2$,
which yield the strength of the coupling between the bare rho
and the two-pion continuum and the four-pion continuum respectively.
Since the two decay channels are both open at these energies, There are
also have
two independent measurements of these two coupling constants;
namely, the two-pion decay and the four-pion decay. 

\begin{figure}
\centerline{\includegraphics[width=9cm]{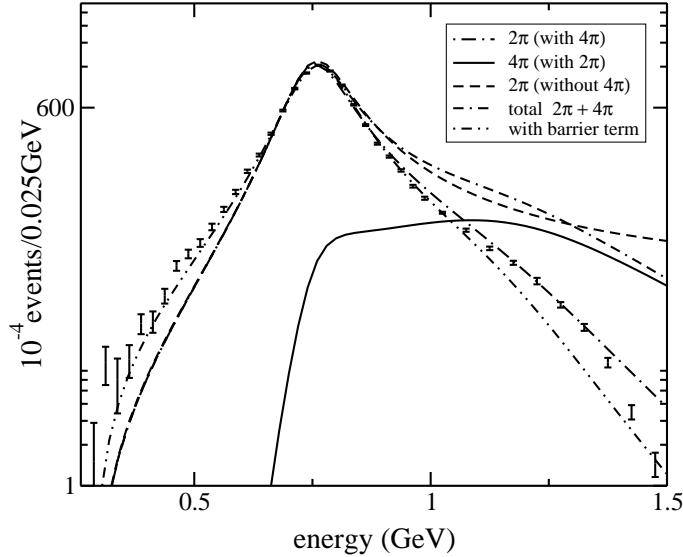}}
\caption{The Fano calculation compared with the $\tau^-$ decay data
from the CLEO collaboration~\cite{CLEO}. There are three cases: the two-pion results
without four-pion contributions (long dashed), the results for the two-pion (dot-dashed) 
and four-pion (solid) continua and their sum (dot-dash-dashed), and the change in the 
two-pion decay of the latter if an artificial barrier term~\cite{CLEO} is introduced, and coupling 
constants refitted (dot-dot-dashed).}
\label{fig4}
\end{figure}

\section{Results}
\label{results}

The hadronic tau decay at the CLEO experiment is 
dominated by the rho meson in the intermediate state.
The initial tau lepton decays into a neutrino and a W-boson.
Subsequently the W-boson produces a hadronic state via the standard
model Lagrangian $Wq\bar q$ vertex. After all the electroweak structure 
has been
resolved, the hadronic part of the decay is dominated by the rho meson 
resonance at
energies between the hadronic decay threshold, which is the two-pion
threshold and about $1$ GeV when the $\omega\pi$ channel is open and
the first of many other decay channels start to compete.

The experimental data from CLEO~\cite{CLEO} is represented in
Fig.~\ref{fig4}, where the number of events, freed from electroweak 
dependences, is plotted as a function
of the invariant mass of the two-pion system. The same figure summarizes
our analysis. If the bare rho state couples only to a two-pion
continuum, the high-energy tail of the resonance
peak is grossly overestimated. 
If the four-pion continuum is included, the two-pion decay channel
is suppressed at higher energies, in favor of four-pion decay (they
cross just above $1$ GeV). The total pionic decay comes again to the
same value as the original two-pion-only result, however, the shape of
the inclusive resonance peak at these higher energies is different. 

For completeness, a fit with a barrier term is also
included~\cite{BW52},
as was used in the data analysis~\cite{CLEO}.
The barrier term has its origin in nuclear quantum mechanics
and relates the amplitude of the asymptotic  wave function
with the inner part in the nuclear potential.
These two parts are matched at the Coulomb barrier through which
an escaping radiative-decay particle tunnels. The momentum dependence
of this matching is reflected in the barrier term. There is no
field-theoretical analogue of the barrier term. It reproduces the right
low-energy behavior; however, within our analysis, which has only two
parameters, we pay for the rather artificial low-energy
improvement with a failure at higher energies. 
The position and the slope of the four-pion decay events compare well with 
both the OPAL~\cite{OPAL} and the CLEO~\cite{CLEO} plots for four-pion
events.

Comparing the resonance shape with the results from our Hamiltonian and 
fitting the two coupling constants $g$ and $g_2$ to hadronic decay of 
the $\tau^-$ we find a number of results.
First, from the fit of the $\pi\pi$ decay of the $\rho$-meson, we find a 
coupling constant
$g_2 \approx 0.2$ to the four-pion state, which agrees well, without any
fitting, with the actual four-pion
data~\cite{OPAL} between threshold ($0.558$ GeV) and $1.1$ GeV. Beyond
$1.1$ GeV we expect other contributions,
which decay to four pions, like $\omega \pi$, to be important.

Second, if we consider the pion as an elementary particle, the initial state, 
as generated by the 
electroweak process, will be the discrete rho state. From the actual
two-pion data which falls of as $\omega^{-2}$ for high energies we infer
that there is no direct coupling to the two-pion continuum,
as this would lead to large amplitudes for large energies. 
Corrections to this assumption of the pion as an asymptotic but pointlike particle,
which follows from combining Fano theory with the chiral Lagrangian, 
could improve the low-energy fit but would fail to descibe the data at higher
energies~\cite{KKW}.

Third, the presence of the four-pion continuum changes the two-pion coupling constant. 
It is interesting to notice that the coupling constants have to be adjusted if additional,
non-interacting channels are included. The competition between decay modes 
generates a simple, but effective final-state interaction.

Fourth, the real part, after the on-shell 
renormalization, yields a negligible effect. 
Since, with on-shell renormalization, the real part vanishes when the energy equals 
the rho mass, the value of the real part is small in the neighborhood of this value.

Finally, we are unable to get a very good fit to the data near threshold, within the model.
There have been several suggestions to improve the
results~\cite{KKW,HLS}, based on chiral symmetry. 
These all use a direct electroweak-pion-pion coupling, which violates 
unitarity for large energies. Physically there are two explanations for the deviations,
within our scheme.
Firstly, the coupling between the electroweak state and the bare rho could have a momentum
dependence, due to, for example, a pionic, or $q\bar q$ quark, rho wave function. 
This should be
visible in an additional quenching of the electroweak decay modes at the energies 
just above the
two-pion threshold. Secondly, the electroweakly produced initial state could contain 
a two-pion state, but only at low energies. It is hard to motivate this from a underlying local
theory. This is equivalent to unitarized hidden local symmetry (HLS)
models~\cite{HLS}.

\section{Discussion}

For the problem of the rho meson studied here we did not require the full machinery
of the Fano theory. For example, there is only one discrete state in the system.
In the case of a neutral rho meson the mixing with the $\omega$ meson could be taken into
account explicitly and unitarily.

For the moment it has been more important to show that the two-pion and the
four-pion continua describe the rho resonance well, where the four-pion
continuum results in two independent observations: its effect on the two-pion
decay, and a direct measurement of the four-pion decay of the rho meson. 
All this can be discussed at the level of {\it restricted Hamiltonians}.
These Hamiltonians contain given states and continua which are expected to 
dominate the results (in this case a composite continuum consisting of 
two- and four-pion states in the rho channel, and the bare rho state).
The fact that no further approximation is required allows one to draw definite
conclusions, whether these states and continua actually dominate the physics
in a certain channel and at a certain energy, or other degrees of
freedom are 
necessary. In this case the data between $0.6$ GeV and $1.1$ GeV is well-described
by these degrees of freedom.

The Fano theory allows one to go further in particle numbers and complexity than 
other methods.  Complex states can be build up hierarchically by admixing states
in their expected order, similar to
angular momentum coupling schemes in atomic physics. The only aspect not handled in
Fano theory is the particle number conserving interaction between continua as occurs
in field theory. Instead of these self-interacting states we assume the states to be 
eigenstates already, with the only noticeable effect a deformation of the phase space 
in $W$, with respect to the free-particles phase space, due to the changed dispersion 
relation $\omega({\bf k})\not = \sum_i \sqrt{{\bf k}^2 + m_i^2}$, to be calculated
separately or approximated by the free result. 

In the future we plan to apply this method to baryonic resonances, 
for which a lot of new data is becoming available from recent JLAB experiments.
In these experiments the problems with data analysis are more stringent due
to the rich spectra and the higher energies~\cite{VDL}.
We will also investigate the way in which to extend Fano theory to allow for
continuum-continuum interactions.

\section*{Acknowledgments}
I would like to thank Wim Ubachs for pointing me in the right direction on
atomic physics, Wolfram Weise for suggesting this problem and the
encouragement, and Rob Timmermans for useful and stimulating
discussions. I would also like to thank Evgeni Kolomeitsev for pointing out
useful references and reading material.
Finally, I gratefully acknowledge the careful reading and extensive
comments by Eric Swanson and Steve Dytman that helped to shape this paper.


\begin{thebibliography}{100}
\bibitem{VDL} T.~P.~Vrana, S.~A.~Dytman, and T.~S.~H. Lee, Phys.\ Rept.\ {\bf 328}, 
181 (2000).
\bibitem{MS59} P.~T.~Matthews and A.~Salam, Phys.\ Rev.\ {\bf 115}, 1079
(1959).
\bibitem{Zwa63} D.~Zwanziger, Phys.\ Rev.\ {\bf 131}, 2818 (1963).
\bibitem{BW52}J.~M.~Blatt and V.~F.~Weisskopf, {\it Theoretical nuclear
physics}, (New York, Wiley, 1952); H.~Feshbach, {\it Theoretical nuclear
physics}, (New York, Wiley, 1991).
\bibitem{tola} S.~Jadach, J.~H.~K\"uhn, and Z.~Was, Comput.\ Phys.
Commun.\ {\bf 64}, 275 (1991); {\bf 70}, 69 (1992); {\bf 76}, 361 (1993).
\bibitem{KKW} F.~Klingl, N.~Kaiser, and W.~Weise, Z.~Phys.~{\bf A 356}, 193 (1996).
\bibitem{HLS} H.~B.~O'Connell {\it et al.}, Nucl.\ Phys.\ {\bf A 623}, 559 (1997); M.~Benayoun
{\it et al.}, Z.\ Phys.\ {\bf C 72}, 221 (1996).
\bibitem{Lut00}M.~F.~Lutz and E.~E.~Kolomeitsev,
Found.\ Phys.\  {\bf 31}, 1671 (2001).
\bibitem{oller}
J.~A.~Oller and E.~Oset,
Phys.\ Rev.\ D {\bf 60}, 074023 (1999).
\bibitem{Pic01}
M.~A.~Pichowsky, A.~Szczepaniak, and J.~T.~Londergan,
Phys.\ Rev.\ D {\bf 64}, 036009 (2001).
\bibitem{tandy} P.~Tandy (private communication); P.~Maris and P.~C.
Tandy, Phys.\ Rev.\ C {\bf 60}, 055214 (1999).
\bibitem{Fan61}U.~Fano, Phys.\ Rev.\ {\bf 124}, 1866 (1961).
\bibitem{Cow82}R.~D.~Cowan, {\it The Theory of Atomic Structure and Spectra},
(University of California Press, 1982).
\bibitem{BR97}S.~M.~Barnett and P.~M.~Radmore, {\it Methods in theoretical quantum optics},
(Oxford Press, Oxford, 1997).
\bibitem{BR69} L.~Banyai and V.~Rittenberg, Phys.\ Rev.\ {\bf 184}, 1903 (1969).
\bibitem{Eck}E.~von Toerne (private communication); K.~W.~Edward {\it et
al.} (CLEO), Phys.\ Rev.\ D {\bf 61}, 072003 (2000);
\bibitem{Sak60} J.~Sakurai, Ann.\ Phys.\ (N.Y.) {\bf 11}, 1 (1960).
\bibitem{BR86} J.~P.~Blaizot and G.~Ripka, {\it Quantum Theory
of Finite Systems}, {(MIT Press, Cambridge (MA), 1986)}.
\bibitem{Lig00}N.~E.~Ligterink, Phys.\ Rev.\ D {\bf 61}, 105010 (2000).
\bibitem{CLEO} S.~Anderson {\it et al.} (CLEO), Phys.\ Rev.\ D {\bf 61}, 112002 (2000).
\bibitem{OPAL} K.~Ackerstaff {\it et al.} (OPAL), Eur.\ Phys.\ J.\ {\bf C 7} 571 (1999) 
\end{thebibliography}
\end{document}